\documentclass{article}
\usepackage{times}
\usepackage{amsmath}
\usepackage{amssymb}

\newcommand{\Tr}{{\mathrm{Tr}}}
\newcommand{\ket}[1]{\left | \, #1 \right \rangle}
\newcommand{\bra}[1]{\left \langle #1 \, \right |}
\newcommand{\proj}[1]{\ket{#1}\!\!\bra{#1}}
\newtheorem{theorem}{Theorem}
\newcommand{\qed}{\hfill $\square$}
\begin{document}

\title{Additivity of the Classical Capacity of Entanglement-Breaking 
Quantum Channels}
\author{Peter W. Shor \\
{\small AT\&T Labs, Florham Park, NJ 07922, USA}}
\maketitle
\abstract{
\em We show that for the tensor product of an entanglement-breaking
quantum channel with an arbitrary quantum channel, both the
minimum entropy of an output of the channel and the 
Holevo-Schumacher-Westmoreland capacity are additive.  In addition, 
for the tensor product of two arbitrary quantum channels, we give a
bound  involving entanglement of formation
for the amount of subadditivity (for minimum entropy output) or
superadditivity (for classical capacity) that can occur.}
\bigskip

One of the more important open questions of quantum information theory
is the determination of the
capacity of a quantum channel for carrying classical information.
This question has been only partially resolved.  If entanglement between 
multiple inputs 
to the channel is not allowed, a formula for the classical capacity of a
quantum channel has indeed been discovered \cite{Hol-cap, SW1}.  This
capacity formula for a quantum channel ${\Psi}$ is
\begin{equation}
\label{def-chi}
\chi^*( {\Psi} ) = \max_{p_i, \rho_i} \ 
H\Big( \sum_i {\Psi} (p_i \rho_i)\Big) - \sum_i p_i H\Big({\Psi}(\rho_i)
\Big)
\end{equation}
where $H$ is the von Neumann entropy $H(\rho) = -\Tr \rho \log \rho$, and 
where the maximization is over probability distributions $p_i$ on density
matrices $\rho_i$ over the input space of the channel.  This 
maximum can be attained because we need at most $d^2$ density 
matrices $\rho_i$ to achieve any attainable value of 
\begin{equation}
\chi(\{p_i, \rho_i\}) =
H\Big( \sum_i {\Psi} (p_i \rho_i)\Big) - \sum_i p_i H\Big({\Psi}(\rho_i)
\Big)
\end{equation}
and are thus maximizing over a compact space.
The general capacity
of a quantum channel ${\Psi}$, without feedback or prior entanglement between
sender and receiver, but possibly using entangled
inputs, is 
\begin{equation}
C({\Psi}) = \lim_{n\rightarrow \infty} 
\frac{1}{n} \chi^*({\Psi}^{\otimes n}),
\end{equation}
i.e., the limit for large $n$ of the capacity when we permit the
input to be entangled over blocks of $n$ channel uses.  This limit
can be shown to
exist because $\chi^*$ satisfies the superadditivity condition
\begin{equation}
\chi^*({\Psi} \otimes {\Phi}) \geq \chi^*({\Psi}) + \chi^*({\Phi}).
\end{equation}
It is conjectured that equality holds, i.e., that
$\chi^*$ is additive, in which 
case $\chi^*$ would give the classical capacity of a quantum channel
without feedback.  Substantial
work has been done on this conjecture \cite{AH, AHW}, and it has been
proven for several special cases.  In particular, it has been proven
when one of the channels is the identity channel \cite{AHW, SW2}, 
when one of the
channels is what A. S. Holevo calls a c-q or q-c channel (these terms will be 
defined later) \cite{Hol-survey, King2001a}, and when one of the channels 
is a unital qubit channel \cite{King2001b}.  

We will prove additivity for the special case where one of the two channels 
is entanglement breaking.  Entanglement breaking channels are channels
which destroy entanglement with other quantum systems.
That is, when the input state is entangled between the input space
${\cal H}_{\mathrm{in}}$ and another quantum system
${\cal H}_{\mathrm{ref}}$, the output of the channel is no longer 
entangled with the system ${\cal H}_{\mathrm{ref}}$.  
Both c-q and q-c channels are special cases of 
entanglement breaking channels.
A c-q channel
is a channel which can be expressed by the composition of a complete 
von Neumann measurement on the input space followed by an arbitrary 
completely positive trace-preserving (CPT) map.  A
q-c channel can be expressed as the composition of
a CPT map followed by a complete
von Neumann measurement on the output space.  Stated more intuitively, 
for c-q maps, the input can be treated as being classical, and for q-c maps, 
the output can be taken to be classical.  In either case, the von Neumann
measurement eliminates any entanglement between the input space and 
another system, so c-q and q-c maps are both special cases of entanglement 
breaking channels.  In a conversation with the author, Michal Horodecki
\cite{Horodecki} gave a simple proof that
any entanglement breaking
channel can be expressed as a \hbox{q-c-q} channel; that is, the 
composition of
a CPT operator followed by a complete von Neumann measurement followed 
by another CPT operator.  (See also \cite{Ruskai} for details of this
proof.) As a consequence, the action
of an entanglement breaking channel $\Phi$  on a state $\rho$ can always
be written in the  following form introduced by Holevo
\cite{Hol-survey}:
\begin{equation}
\label{Holevo-channel}
\Phi( \rho) = \sum_{i} \, \Tr \big(X_i \rho \big) \, {\theta}_i
\end{equation}
where $\{ X_i \}$ form a general POVM and $\{ {\theta}_i \}$ are
arbitrary states. For a c-q map, $ X_i  = \proj{i}$ where $\ket{i}$ form
an orthonormal basis, and for a q-c map ${\theta}_i = \proj{i}$.

The additivity problem for capacity is closely related to another additivity
problem; that of the minimum entropy output of a channel \cite{KR}.  
For the case
of entanglement breaking channels, we first found the additivity proof for 
the minimum entropy output, and then discovered a straightforward way to
extend this additivity proof to cover the classical capacity.  In this paper, 
we first give the proof
for additivity of minimum entropy output, as this proof contains the 
important ideas for the capacity proof, but has
significantly fewer technicalities.
\begin{theorem}
\label{additivity-minent}
For an arbitrary quantum channel $\Psi$, and an
entanglement breaking channel ${\Phi}$
\begin{equation}
\min_{\rho_{AB}} H \Big(( \Psi \otimes \Phi )(\rho_{AB})\Big) 
= \min_{\rho_A}  
H\Big( \Psi(\rho_A) \Big) +\min_{\rho_B} H \Big( \Phi (\rho_B)\Big) .
\end{equation}
\end{theorem}
{\bf Proof:}
The left-hand side is clearly at most the right-hand side, as can be seen
by choosing $\rho_{AB} = \rho_A \otimes \rho_B$.  We would
like to show that it is at least the right-hand side.
We use the strong subadditivity property of von Neumann entropy \cite{LR}.  
Consider the minimum obtainable value of
$H\big((\Psi \otimes \Phi)(\rho_{AB})\big)$.  
Because $\Phi$ is entanglement breaking,
\begin{equation}
(I\otimes \Phi)(\rho_{AB}) = \sum_j q_j \proj{a_j} \otimes \proj{b_j}
\end{equation}
for some $q_j$, $\ket{a_j}  \in {\cal H}_A$ and $\ket{b_j}  \in {\cal H}_B$.
Now, we apply to the state
\begin{equation}
\sigma_{ABC} = \sum_j q_j \Psi(\proj{a_j}) \otimes \proj{b_j} \otimes \proj{j}
\label{sigmaABCdef}
\end{equation}
the property of strong subadditivity in the form
\begin{equation}
H(\sigma_{AB}) \geq H(\sigma_{ABC}) - H(\sigma_{BC}) + H(\sigma_B).
\end{equation}
We have
\begin{eqnarray}
\label{sigmaAB}
\sigma_{AB} &=&
\sum_j  q_j \Psi(\proj{a_j}) \otimes \proj{b_j} \\
&=& (\Psi \otimes \Phi)(\rho_{AB}), \nonumber
\end{eqnarray}
the quantity for the entropy of which we would like a 
lower bound.
Now, note that
\begin{eqnarray}
H(\sigma_{ABC}) - H(\sigma_{BC}) &=& H(\sigma_{AC}) - H(\sigma_C) \\
        &=& \sum_j q_j H\Big(\Psi(\proj{a_j})\Big). \nonumber
\end{eqnarray}
The first equality above follows from the facts that, 
in Eq.~(\ref{sigmaABCdef}), the $\proj{j}$ form
an orthonormal set and the $\proj{b_j}$ are pure
states, so that $H(\sigma_{ABC}) = H(\sigma_{AC})$ and 
$H(\sigma_{BC}) = H(\sigma_C)$.  The second equality follows from the 
chain rule for entropy,  namely
\begin{equation}
\label{chain}
H\Big(\sum_j q_j \gamma_j \otimes \proj{j} \Big) - 
\sum_j H\Big(q_j \proj{j} \Big) = \sum_j q_j H( \gamma_j) .
\end{equation}
Now, note that
\begin{eqnarray}
\label{sigma-B}
\sigma_B &=& \sum_j q_j \proj{b_j} \\ \nonumber
&=& \Tr_A (I \otimes \Phi)(\rho_{AB})\\ \nonumber
&=& \Phi(\Tr_A \rho_{AB})
\end{eqnarray}

Putting the above equalities together, we see that
\begin{equation}
H\Big((\Psi \otimes \Phi)(\rho_{AB})\Big) 
\geq \sum_j q_j H\Big(\Psi(\proj{a_j})\Big) 
+ H\Big(\Phi(\Tr_A\rho_{AB})\Big).
\end{equation}
Since $\sum_j q_j =1$, the
right-hand side is clearly at least the sum of the minimum output entropies of
$\Psi$ and of $\Phi$.  We have thus shown that
the minimum output entropy is additive for the tensor product of two
channels if one of the channels is an entanglement breaking channel.
\qed

We now prove the corresponding additivity result for the 
Holevo-Schumacher-Westmoreland capacity $\chi^*$; recall  
\begin{equation}
\chi^*({\Psi}) = \max_{p_i, \rho_i} H\Big({\Psi}(\sum_i p_i \rho_i) \Big)
- \sum_i p_i H\Big({\Psi}(\rho_i)\Big)
\end{equation}
over probability distributions $p_i$ and density
matrices $\rho_i$.  
\newpage

\begin{theorem}
\label{additivity-chi}
For an arbitrary quantum channel $\Psi$, and an
entanglement breaking channel ${\Phi}$
\begin{equation}
\chi^* ( \Psi \otimes \Phi )
=  \chi^*( \Psi ) +\chi^* ( \Phi )
\end{equation}
\end{theorem}
{\bf Proof:}
The capacity $\chi^*$ is composed of two terms.  We will be treating these 
two terms separately.  For the second term, additivity is shown in
essentially the same way as in the proof of additivity for minimum entropy,
and for the first term, additivity follows from the subadditivity
of von Neumann entropy.  

Again, we assume that we have an arbitrary quantum channel $\Psi$, 
and an entanglement breaking channel $\Phi$.
We use strong subadditivity.  Consider the optimal signal states
for $\Psi \otimes \Phi$, i.e., the $p_i$
and $\rho_i$ such that
\begin{equation}
\chi^* ( \Psi \otimes \Phi) = H\big( (\Psi \otimes \Phi ) ( \rho) \big)
- \sum p_i H\big( ( \Psi \otimes \Phi) (\rho_i)\big)
\end{equation}
where $\rho = \sum p_i \rho_i$.  Let us consider the state
$(I \otimes \Phi)(\rho_i)$.  Because $\Phi$ is an
entanglement breaking map, this state is separable, and so
\begin{equation}
(I \otimes \Phi) (\rho_i) = 
\sum_j q_{ij} \proj{a_{ij}} \otimes \proj{b_{ij}}
\end{equation}
for some $q_{ij}, \ket{a_{ij}}$, $\ket{b_{ij}}$.
Now, we apply strong subadditivity to the state 
\begin{equation}
\sigma_{ABC} = 
\sum_j q_{ij} \Psi ( \proj{a_{ij}} ) \otimes \proj{b_{ij}} \otimes \proj{j}.
\end{equation}
To simplify notation, we let the dependence of $\sigma$ on $i$ be implicit.
Again, we apply strong subadditivity in the form
\begin{equation}
H(\sigma_{AB}) \geq H(\sigma_{ABC}) -H(\sigma_{BC}) + H(\sigma_{B})
\end{equation}
As before, 
\begin{equation}
H(\sigma_{AB}) = H\Big( (\Psi \otimes \Phi )(\rho_i)\Big).
\end{equation}
We also have that
\begin{equation}
H(\sigma_{B}) = H\Big(\Phi(\Tr_A \rho_i)\Big)
\end{equation}
and
\begin{equation}
H(\sigma_{ABC}) - H(\sigma_{BC}) 
= \sum_j q_{ij} H\Big(\Psi( \proj{a_{ij}} ) \Big).
\end{equation}
We let $\proj{a_{ij}} = \tau_{ij}$. 
Then $\Tr_B \rho_i = \sum q_{ij} \tau_{ij}$.
Combining the terms, we observe 
\begin{equation}
H\Big( (\Psi \otimes \Phi )(\rho_i)\Big) \geq 
\sum_j q_{ij} H\Big(\Psi(\tau_{ij})\Big) + H\Big(\Phi(\Tr_A\rho_i)\Big)
\end{equation}
Now, let us sum over all the states $\rho_i$.  We obtain
\begin{equation}
\label{second-term}
\sum_i p_i H\Big( (\Psi \otimes \Phi )(\rho_i)\Big) \geq 
\sum_{i,j} p_i q_{ij} H\Big(\Psi(\tau_{ij})\Big)+ \sum_i p_i H\Big(\Phi(\Tr_A\rho_i)\Big).
\end{equation}
Using subadditivity of von Neumann entropy and the above inequality
(\ref{second-term}), we get
that
\begin{eqnarray}
\label{thm2-eq}
\chi^* ( \Psi \otimes \Phi) &=&  
H\Big((\Psi \otimes \Phi)(\rho)\Big) 
- \sum_i p_i H\Big( (\Psi \otimes \Phi )(\rho_i)\Big) \\
\nonumber
&\leq & H\Big(\Psi(\Tr_B \rho)\Big) + H\Big(\Phi(\Tr_A \rho)\Big)  
\\&& \qquad
- \sum_{i,j} p_i q_{ij} H\Big(\Psi(\tau_{ij})\Big) 
- \sum_i p_i H\Big(\Phi(\Tr_A\rho_i)\Big).
\nonumber
\end{eqnarray}
However, since 
\begin{equation}
\sum_{i,j} p_i q_{ij} \tau_{ij} = \sum_i p_i \Tr_B \rho_i =
\Tr_B \rho \qquad {\mathrm{and}} \qquad
\sum_i p_i \Tr_A \rho_i = \Tr_A \rho,
\end{equation}
we see that
\begin{equation}
\chi^* ( \Psi \otimes \Phi) \leq \chi^* (\Psi) + \chi^*(\Phi).
\end{equation}
As the opposite inequality is easy, we have additivity of $\chi^*$ for
entanglement breaking channels.
\qed

We finally give a bound on the amount of superadditivity for general
channels.  For this,
we need to define the entanglement of formation of a bipartite state.
This is another quantity that is conjectured
to be additive, but for which additivity has not been proved. 
Entanglement of formation for a bipartite state $\rho_{AB}$ is defined 
\begin{equation}
E_F(\rho_{AB}) = \min_{{p_i, \rho_i}\atop{\scriptstyle
\sum_i p_i \rho_i = \rho_{AB}}}
\sum p_i H(\Tr_A \rho_{i})
\end{equation}
where the minimization is over probability distributions $p_i$ on rank-one 
density matrices $\rho_i$ such that $\sum_i p_i \rho_i = \rho_{AB}$.  
The theorem is
\begin{theorem}
\label{quant-superadditivity}
Suppose we have two quantum channels, i.e., completely positive
trace preserving maps, $\Psi$ and $\Phi$.  Then
\begin{eqnarray}
\min_{\rho_{AB}} H\Big(({\Psi\otimes \Phi})(\rho_{AB})\Big)  
\ \geq\  \min_{\rho_A} H\Big(\Psi(\rho_A)\Big)
\hspace{-1ex} & + &\hspace{-1ex}  \min_{\rho_B}H\Big(\Phi(\rho_B)\Big) \\
\hspace{-1ex} \nonumber
& - & \hspace{-1ex} \max_{\rho_{AB}} E_F\Big((I\otimes \Phi)(\rho_{AB})\Big)
\end{eqnarray}
and
\begin{equation}
\label{chi-bound}
\chi^*({\Psi\otimes \Phi})  \leq \chi^*(\Psi)+ \chi^*(\Phi)
+ \max_{\rho_{AB}} E_F\Big((I\otimes \Phi)(\rho_{AB})\Big).
\end{equation}
\end{theorem}
Note that the formulation of the theorem is asymmetric with respect to $\Psi$ and
$\Phi$.  Thus, to bound the amount of sub- or superadditivity, 
one can use either the entanglement of formation of 
$(I\otimes \Phi) (\rho_{AB})$ or of $(\Psi\otimes I) (\rho_{AB})$, whichever 
is smaller.

\noindent{\bf Proof}:
We first give the proof of the first part 
of Theorem~\ref{quant-superadditivity}.   
Let 
\begin{equation}
\label{decompose}
(I \otimes \Phi) (\rho_{AB}) = \sum_i q_i \nu_i
\end{equation}
be the decomposition of $(I \otimes \Phi)(\rho_{AB})$ into pure 
states $\nu_i$ that minimizes entanglement of formation, i.e., so that 
$\sum_j q_j H(\Tr_A \nu_j)$ is minimum.  Now, we consider
\begin{equation}
\sigma_{ABC}  = \sum_j q_j (\Psi \otimes I) (\nu_j ) \otimes \proj{j}
\end{equation}
and apply strong subadditivity to this state.  
We obtain
\begin{equation}
\label{strong-sub-C}
H(\sigma_{AB}) \geq \Big( H(\sigma_{ABC}) - H(\sigma_C)\Big) + 
\Big( H(\sigma_B) \Big) - \Big( H(\sigma_{BC}) - H(\sigma_C) \Big).
\end{equation}
As in (\ref{sigmaAB}), we have
\begin{equation}
H(\sigma_{AB}) = H\Big( (\Psi\otimes\Phi)(\rho_{AB})\Big).
\end{equation}
Similar to (\ref{sigma-B}), we get
\begin{equation}
\label{eq1}
H(\sigma_B) = H\Big(\sum_j q_j \Tr_A \nu_j\Big) =
H\Big(\Phi(\Tr_A \rho_{AB}) \Big)
\end{equation}
Furthermore, the choice of $\nu_j$ and the definition of $E_F$ gives
\begin{equation}
\label{eq2}
H(\sigma_{BC}) - H(\sigma_C) = E_F \Big( (I \otimes \Phi)(\rho_{AB})\Big)
\end{equation}
Finally application of the entropy chain rule (\ref{chain}) gives
\begin{equation}
\label{eq3}
H(\sigma_{ABC}) - H(\sigma_C) = \sum_{j} q_j H\Big((\Psi \otimes I)(\nu_j)\Big)
\end{equation}
The expression (\ref{eq1}) is bounded below by
$\min_{\rho}H\big(\Phi(\rho)\big)$. The second expression (\ref{eq2}) is bounded
above by $\max_{\rho} E_F \big( (I \otimes \Phi)(\rho)\big)$.
The third expression (\ref{eq3}) is bounded below by
$\min_{\rho} H\big((\Psi \otimes I)(\rho)\big)$, which is known to
equal $\min_{\rho} H\big(\Psi(\rho)\big)$. Combining these three
expressions give the first part of 
Theorem \ref{quant-superadditivity}. 

To prove the second part of the theorem, (\ref{eq3}) must be
replaced by
\begin{equation}
\label{ineq3}
H(\sigma_{ABC}) - H(\sigma_C) \geq \sum_{jk} q_j r_{jk} 
H\Big(\Psi(\proj{v_{jk}})\Big)
\end{equation}
for states $\ket{v_{jk}}$ and probabilities $q_j r_{jk}$ such
that
\begin{equation}
\label{new-sum}
\sum_{j,k} q_j r_{jk} \proj{v_{jk}} = \Tr_B \nu_j.
\end{equation}
We then consider the signal states $\rho_i$ and the associated
probabilities $p_i$
which give the value of $\chi^*(\Psi \otimes \Phi)$ in 
Equation~(\ref{def-chi}), and
let $\sum_i p_i \rho_i = \rho$.  We now
use expressions (\ref{eq1}), (\ref{eq2}), (\ref{ineq3}) with
$\rho_i$ in the place of $\rho_{AB}$.  Combining these three expressions
yields
\begin{eqnarray}
H\Big( (\Psi \otimes \Phi)(\rho_i)\Big) \ \geq \ 
H\Big(\Phi(\Tr_A \rho_i)\Big) \hspace*{-1ex}  
&+& \hspace*{-1ex} \sum_{j,k} q_{ij} r_{ijk} H\Big(\Psi(\proj{v_{ijk}})\Big)\\
&-& \hspace*{-1ex} E_F\Big( (I\otimes \Phi)(\rho_i)\Big). \nonumber
\end{eqnarray}
The second part of Theorem \ref{quant-superadditivity} then follows 
in a way entirely analogous to the proof of Theorem \ref{additivity-chi}.
We use the equalities
\begin{equation}
\Tr_B \rho = \sum_i p_i\Tr_A \rho_i 
\end{equation}
and 
\begin{equation}
\Tr_A \rho = \sum_{i,j,k} p_i q_{ij} r_{ijk} \proj{v_{ijk}},
\end{equation}
and expand $\chi^*(\Psi \otimes \Phi)$ similarly to Eq.~(\ref{thm2-eq}) 
to obtain Eq.~(\ref{chi-bound}).
 
We still must prove the inequality (\ref{ineq3}).  
The left hand side of (\ref{ineq3}) is
\begin{equation}
\sum_j {q_j} H\Big( (\Psi\otimes I)(\nu_j) \Big) 
\end{equation}
Now, $\nu_j$ is a purification of $\Tr_B \nu_j$, and 
$H\big( (\Psi\otimes I)(\nu_j) \big) = H\big( (\Psi \otimes I) (\tau)\big)$ 
for any other quantum state $\tau$ which is a purification of 
$\sigma_j = \Tr_B \nu_j$.  Let $\sigma_j = \sum_k q_{jk} \proj{v_{jk}}$ be the 
eigenvector decomposition of $\sigma_j$.  A different purification is
\begin{equation}
\tau_j = \Big( \sum_k q_{jk} \ket{v_{jk}} \otimes \ket{k}\otimes \ket{k} \Big)
\Big(\sum_k q_{jk} \bra{v_{jk}} \otimes \bra{k}\otimes \bra{k} \Big)
\end{equation}
It suffices to show that 
\begin{equation}
H\Big( (\Psi \otimes I) (\tau_j)\Big) \geq
H\Big( \Tr_{3}(\Psi \otimes I) (\tau_j)\Big) -
H\Big( \Tr_{12}(\Psi \otimes I) (\tau_j)\Big) 
\end{equation}
as the first term in the above equation is $H\big( (\Psi\otimes I)(\nu_j) 
\big)$, the second is $H(\{q_{jk}\}_k) + 
\sum_k q_{jk} H\big(\Psi(\proj{v_{jk}})\big)$, and the 
third is $H(\{q_{jk}\}_k)$.  However, the above equation follows from
the inequality
$H(\rho_{34}) \geq H(\rho_3) - H(\rho_4)$, which is a consquence
(after another purification)
of the subadditivity property of entropy \cite{AkLb}.
\qed

\section*{Acknowledgements}
I would like to thank Michal and Pawel Horodecki, Chris King,
and Beth Ruskai for interesting and informative conversations.  In 
addition, I would like to thank Michal Horodecki for discovering the proof 
that all entanglement-breaking channels can be expressed in Holevo's form 
(i.e., Eq.~(\ref{Holevo-channel})), and Chris King and Beth Ruskai for 
their extremely helpful suggestions for improving the exposition of 
this paper.

\end{document}